\documentclass[floatfix,superscriptaddress,pra, amsmath,amssymb,notitlepage,nofootinbib,reprint]{revtex4-1}
\usepackage{graphicx}
\usepackage{dcolumn}
\usepackage{bm}
\usepackage{hyperref}

\usepackage{amsmath}
\usepackage{amsthm}

\usepackage{amsfonts,dsfont}
\usepackage{graphicx}
\usepackage{bbm}
\usepackage{color}
\usepackage{comment}

\theoremstyle{remark}
\newtheorem{remark}{Remark}

\theoremstyle{theorem}
\newtheorem*{proposition}{Proposition}

\newcommand{\bra}[1]{\langle #1 |} 
\newcommand{\ket}[1]{| #1 \rangle } 
\newcommand{\tr}[0]{\mathrm{tr}} 
\newcommand{\upd}[0]{\mathrm{d}}

\newcommand{\re}{\mathfrak{R}\mathrm{e}}

\definecolor{cbl}{rgb}{0,0,1}

\definecolor{crd}{rgb}{1,0,0}

\begin{document}


\title{Computing the Rates of Measurement-Induced Quantum Jumps}

\author{Michel Bauer}
\email{michel.bauer@cea.fr}
\affiliation{
Institut de Physique Th\'eorique, CEA Saclay and CNRS, Gif-sur-Yvette, France
}
\author{Denis Bernard}
\email{denis.bernard@ens.fr}
\author{Antoine Tilloy}
\email{antoine.tilloy@ens.fr}
\affiliation{Laboratoire de Physique Th\'eorique, CNRS and Ecole Normale Sup\'erieure de Paris, France
}

\date{\today}
\begin{abstract}
Small quantum systems can now be continuously monitored experimentally which allows for the reconstruction of quantum trajectories. A {peculiar} feature of these trajectories is the emergence of jumps between the eigenstates of the observable which is measured. Using the Stochastic Master Equation (SME) formalism for continuous quantum measurements, we show that the density matrix of a system indeed shows a jumpy behavior when it is subjected to a tight measurement (even if the noise in the SME is \emph{Gaussian}). We are able to compute the jump rates analytically for \emph{any} system evolution, { i.e. any Lindbladian, and we illustrate} how our general recipe can be applied to two simple examples.  {We then discuss} the mathematical, foundational and practical applications of our results. The analysis we present is based on a study of the strong noise limit of a class of stochastic differential equations (the SME) and as such the method may be applicable to other physical situations in which a strong noise limit plays a role.
\end{abstract}

\pacs{Valid PACS appear here}
\maketitle

\section{Introduction}

Recent advances in experimental techniques now allow for a tight monitoring, \textit{i.e.} a continuous and strong measurement, of small open quantum systems. The corresponding quantum trajectories can now be recorded with increasing precision \cite{murch2013,weber2014}. A striking feature of such systems undergoing continuous measurement is the emergence of a jumpy behaviour between measurement  eigenstates. This interesting and ubiquitous phenomenon has already been observed in many experiments \cite{nagourney,bergquist,sauter}. Even if quantum jumps were already well known to Bohr \cite{bohr1913}, to our knowledge, their emergence and statistics have never been studied thoroughly {in the general case.} This theoretical investigation is the main purpose of this article.

We conduct our study using the formalism of continuous measurement developed in \cite{barchielli1986,caves1987,barchielli1991,wiseman1996,belavkin1992,barchielli2009}, that is we study a stochastic differential equation (SDE) \emph{with Gaussian noise} describing the (continuous) evolution of the density matrix of a small open system of interest. The stochasticity comes from the conditioning of the density matrix on the (random) measurement outcomes. Such equations can be obtained as the limit of a series of weak measurements carried out on a quantum system \cite{attal2006,pellegrini2008,pellegrini2009}. In this setting, the jumpy regime arises when the rate of measurements, that we call $\gamma^2$, is large. At this point we should emphasize that these quantum jumps obtained in the large $\gamma$ limit of \emph{continuous} equations are  \emph{not} the same as  those emerging from the intrinsically discontinuous Poissonian unravelling of a quantum master equation. 

From a mathematical perspective, we study a class of non-linear SDE in the
strong noise limit, and show that the solution converges, in a very weak sense,
to a continuous-time finite-state Markov process on the measurement pointer
states and we compute the transition rates. We
show that the finite dimensional distributions converge weakly
towards those of a finite-state Markov process on the measurement pointer
states\footnote{Weak convergence is also called convergence in law: expectations of bounded continuous functions depending on the positions at fixed times $t_1,...,t_k$ have a large $\gamma$ limit which is the expectation with respect to a finite-state Markov process.}. We should stress that the 
  convergence is weak also in the sense that some interesting quantum
  fluctuations, which should be further studied, are preserved in the limit. 

Eventually, we believe this study provides a quantitative understanding on the semi-classical behaviour of tightly monitored quantum systems and heavily generalises the specific cases treated in \cite{jordan2013,oqrw,lmp}. As such it could have applications to a wide class of microscopic open systems showing a jumpy behaviour ranging from quantum dots to photons in a cavity.

\paragraph*{Outline}
This paper is structured as follows. In section \ref{sec:main-res} we present our model and the main claim of the article without proof. We discuss the implications of the result and then study two simple examples of applications in section \ref{sec:applications}. Section \ref{sec:details} is more technical and devoted to the proofs. We first show the emergence of the jumps and compute their rate assuming a given scaling limit. We then proceed to show that this scaling is actually the most general. Eventually, we discuss in more details the physical meaning of our results in section \ref{sec:discussion}.

\section{Main Results} \label{sec:main-res}
We consider a very general quantum system, but with a finite dimensional Hilbert space, whose dynamics are prescribed by a Lindbladian $\mathcal{L}$. We assume that an observable $\mathcal{O}$ is also continuously measured at a rate, or strength, $\gamma^2$ (say with a repeated interaction scheme as in \cite{guerlin2007,hume2007}) with efficiency $\eta$. As a result the density matrix of the system evolves in the following way \cite{wiseman2009}:
\begin{equation}
\label{eq:main}
d\rho_t=\mathcal{L}(\rho_t)\,dt+\gamma^2 L_{N}(\rho_t)\, dt + \gamma \sqrt{\eta}\,{D}_{N} (\rho_t)\, dW_t,
\end{equation}
where $W_t$ is a standard Wiener process, and $N$ is the so-called measurement operator with ${\cal O}=N+N^\dagger$ the measured observable, 
$L_{N}(\rho)=N\rho N^\dagger - \frac{1}{2}\lbrace N^\dagger N,\rho\rbrace$ is the Lindblad generator associated to $N$ and ${D}_{N}(\rho) =N\rho+\rho N^\dagger-\rho\,\tr({\cal O}\rho)$ the stochastic innovation term~\footnote{Notice that we use the same notation as in \cite{epl} for consistency but that the latter differs from Milburn and Wiseman's \cite{wiseman2009}. The dictionary is the following: $L_{N}(\rho)=D[N]\rho$ and ${D}_N(\rho)= \mathcal{H}[N]\rho$. We prefer to use the letter $L$ for the Lindbladian and to reserve the letter $H$ for an Hamiltonian, and we use the letter $D$ for the term multiplying the Brownian noise as a reference to a non-linear diffusion coefficient.}. 
Any given realisation of the Wiener process corresponds to a sample of a time series of measurements. Measurement outputs $x_t$ are random according to the rules of Quantum Mechanics and given by $dx_t= \gamma \tr({\cal O}\rho)\, dt + \eta^{-1/2}\,dW_t$~\cite{wiseman2009}. Solutions of eq.(\ref{eq:main}) are called quantum trajectories. 
We will write everything in the basis where $\mathcal{O}$ is diagonal, i.e.
$\mathcal{O}=\sum_k \lambda_k \ket{k}\bra{k}$ and suppose that all its
eigenvalues are different. We assume that the measurement operators $N$ are
diagonal in this basis, $N=\sum_k \nu_k \ket{k}\bra{k}$ with
$\lambda_k=\nu_k+\bar \nu_k=2\,\re\, \nu_k$, in order to ensure for the process
to be a non-demolition measurement in absence of the dynamics generated by the
Lindbladian ${\cal L}$. The eigenstates $\ket{k}$ will be called pointer states
in what follows.

When $\gamma$ is large, the system density matrix will undergo {\it quantum
  jumps} between the pointer states of the observable $\mathcal{O}$. Our
objective is to characterise those jumps at the stochastic process level and not
only at the ensemble average level, i.e. we want to show that the conditioned
density matrix becomes itself, as far as the diagonal is concerned and in the large $\gamma$ limit, a finite state
Markov process (and not only that the diagonal of
the unconditioned
density matrix is the probability density associated to a finite state Markov
process as in \cite{lesanovsky2013,everest2014}).  Especially, the
objective of this paper is to show how the jump rates between different states
depend on the parameters of the Lindbladian $\mathcal{L}$ and as a result how
they also partially characterise it. 

We first need to say a brief word about the scaling limit in order to state the results, and it will be more carefully explained later {in Section \ref{proof:scaling}}. It is well known that if $\mathcal{L}$ is generated by a simple Hamiltonian, a continuous strong measurement will tend to Zeno freeze the system in one of the pointer states for an arbitrary long time, i.e. when $\gamma\rightarrow \infty$ all the jump rates will go to 0. As a result and to get meaningful predictions in this limit, we need to adequately rescale the different parts of the dynamics to keep finite jump rates in the large $\gamma$ limit. Such a rescaling is not required for all parts of the dynamics because as was argued in \cite{lmp}, jumps that emerge from a dissipative coupling cannot be Zeno frozen. To get the most general scaling limit, we consequently need to split the Lindbladian into different parts, actually four, that need to be rescaled separately. We write $Q_i$ for the diagonal coefficients of $\rho$ in the measurement eigenbasis, the \emph{probabilities}, and $U_{ij}$ for the non diagonal coefficients of $\rho$, the (not yet rescaled) \emph{phases},
\[ Q_i:= \bra{i}\rho\ket{i},\quad U_{ij}:= \bra{i}\rho\ket{j}, \; i\neq j.\]
We decompose $\mathcal{L}$ in four super-operators, $A$ that sends the probabilities to the probabilities, $B$ the phases to the probabilities, $C$ the probabilities to the phases and $D$ the phases to the phases. 
\[
\partial_t\rho_t=\mathcal{L}(\rho_t) \;\underset{\mathrm{notation}}{\Longleftrightarrow} \;\left\lbrace
\begin{array}{c}
\partial_t Q_i= A(\mathbf{Q})_i+B(\mathbf{U})_i\\ 
\partial_t U_{ij}=C(\mathbf{Q})_{ij}+D(\mathbf{U})_{ij}
\end{array}
\right. ,
\]
with $A(\mathbf{Q})_i=A^k_iQ_k$, $B(\mathbf{U})_i=B^{kl}_i U_{kl}$, $C(\mathbf{Q})_{ij}=C^{k}_{ij}Q_k$, and $D(\mathbf{U})=D^{kl}_{ij} U_{kl}$;  summation over repeated indices is implicit.
The reason why this decomposition is legitimate will be clearer later but a good rationale for it is that as the strong measurement will tend to shrink the phases, they will obviously need a differentiated treatment from the probabilities. We now claim that $A$ needs no rescaling, that $C$ and $B$ need to scale like $\gamma$ and $D$ like $\gamma^2$. In what follow, we thus write :
\begin{equation}
A=\mathcal{A},\ B=\gamma \mathcal{B},\ C=\gamma\mathcal{C}, D=\gamma^2\mathcal{D}.
\label{eq:liouvillianscaling}
\end{equation}
For such a scaling to be consistent with the complete positivity of the map generated by $\mathcal{L}$ in the large $\gamma$ limit, we will see that
$\mathcal{D}$ needs to be diagonal:
\[
\mathcal{D}_{ij}^{kl}=-d_{ij}\, \delta^k_i\delta^l_j
\]
We should also add that equation (\ref{eq:liouvillianscaling}) only gives the dominant terms in an expansion in power of $\gamma$ and that the sub-leading corrections may in general be needed for compatibility with the complete positivity of the map associated to $\mathcal{L}$. We just claim that they have no impact on the large $\gamma$ limit as expected and omit them for clarity.

Our main result, which will be proved in section \ref{proof1}, can then be stated as follows:

\begin{proposition} \label{prop:convmark}
With the previous notations, when $\gamma\rightarrow\infty$ the finite dimensional distributions of the conditioned
density matrix $\rho_t$ converge to those of a finite state Markov process on the projectors associated to the measurement eigenvectors. The jump rate from site $i$ to site $j$ then reads in terms of the rescaled coefficients:
\begin{equation} \label{m-rate}
m^{i}_{j}=\mathcal{A}^i_j + 2\, \re\, \sum_{k< l} \frac{\mathcal{C}^i_{kl}\mathcal{B}^{kl}_{j}}{\Delta_{kl}}
\end{equation}
with $\Delta_{kl}:=\frac{1}{2} (|\nu_k|^2+ |\nu_l|^2 -2\nu_k \bar \nu_l)+d_{kl}$.

\end{proposition}

In other words, in the strong measurement limit, the density matrix behaves as a jump process between the projectors $\ket{i}\bra{i}$  with jump rates given by the previous formula. Let us now make several remarks. 
\begin{remark}
The result does not depend on the efficiency $\eta$ of the measurement (provided it does not vanish).
\end{remark}

\begin{remark}
\label{rem:average}
The mean probabilities $\overline{Q}_i:=\mathbb{E}[Q_i]$ obtained by averaging
over quantum trajectories, satisfy the finite state Markov process  equation 
\begin{equation}
\partial_t \overline{Q}_j = \sum_i \overline{Q}_i\, m^i_j.
\label{eq:average}
\end{equation}
\end{remark}

As a result, our framework can also be applied to systems with strong dissipation or equivalently systems that are strongly measured with unrecorded outcomes. In such a situation, the density matrix is diagonal and  its evolution is simply given by the average over trajectories of the jump process~: $\partial_t \overline{Q}_j = \sum_i \overline{Q}_i\, m^i_j$. In that case, the density matrix itself is not a finite state Markov process but the probability distribution of a finite state Markov process. In this simpler setting, result (\ref{eq:average}) can admittedly be derived from the Lindblad equation (\textit{i.e.} the SME averaged over the noise). However, remark \ref{rem:average} shows that it can be seen as a trivial byproduct of our more general proposition.


\begin{remark}\label{rem}
The reader may wonder how our result written in terms of $\mathcal{A}, \mathcal{B}, \mathcal{C}$ and $\mathcal{D}$ may be related to the generators of the Lindbladian. {We} claim and will prove {in Section \ref{proof:scaling}} that the most general scaling that can be written is the following:
\begin{equation}
\begin{split}
\mathcal{L}(\rho)=&-i[\gamma H + \gamma^2 H^{diag},\rho] + \sum_a L_{N_a}(\rho) \\&+\gamma^2 \sum_b L_{N^{diag}_b}(\rho) + \text{subleading terms}
\end{split}
\label{eq:generatorscaling}
\end{equation}
where $L_N$ denotes as above the Lindblad generator associated to $N$ and where the superscript \textit{"diag"} means that the corresponding matrix has to be diagonal, $H$ (without superscript) is any self-adjoint matrix and the $N_a$ (without superscript) can be any matrix. The subleading terms in $\gamma$ are irrelevant for the jump rate computation. Notice that the most general scaling is again far from trivial, possibly with terms of order 0, 1 and 2 in $\gamma$. Using the notation in equation (\ref{eq:generatorscaling}) we get the following expression for the terms appearing in the jump rates in (\ref{m-rate}):
\begin{equation}
\begin{split}
\mathcal{A}^i_j&=\sum_a\Big( \big\vert(N_a)_{ji}\big\vert^2-\delta_{ij} (N_a^{\dagger} N_a)_{jj}\Big)\\
\mathcal{C}^i_{kl}\mathcal{B}^{kl}_{j}&=\left(H_{il}\delta_{ik}-H_{ki}\delta_{il}\right)\left(H_{jk}\delta_{jl}-H_{lj}\delta_{jk}\right)\\
d_{ij}&=\frac{1}{2}\sum_a\Big\lbrace|(N_a^{diag})_{ii}|^2+|(N_a^{diag})_{jj}|^2\\&-2(N_a^{diag})_{ii} (\overline{N_a^{diag})}_{jj}\Big\rbrace+i (H^{diag}_{ii}-H^{diag}_{jj})
\end{split}
\end{equation}
where the second line is understood without summation on repeated indices. If we forget about the terms that need to be rescaled in $\gamma^2$, this means that the main contribution to the jump rates comes from the non diagonal part of the matrices appearing in the Lindblad generators. The second contribution comes from the non diagonal parts of the Hamiltonian that need to be rescaled with a factor $\gamma$ to remain relevant in the strong measurement limit. Before going to the proof of this result, we give two simple examples of application of our formula.

\end{remark}

\section{Applications} \label{sec:applications}
We now study two examples where the jump rates depend in a very different way on the system dynamics. In the first example, jumps will emerge from the competition between a unitary evolution and continuous quantum measurement ($\mathcal{B}\neq 0$ and $\mathcal{C}\neq 0$): when the Hamiltonian is kept constant, the jump rates will go to zero when $\gamma$ goes to infinity, the evolution will be progressively \emph{Zeno-frozen}. In the second example, the jumps will emerge from the competition between a dissipative evolution and continuous quantum measurement ($\mathcal{A}\neq 0$): the jump rates will converge to a constant when $\gamma$ goes to infinity, the evolution will not be \emph{Zeno-frozen}. A system in which those {two kinds of evolution} are present at the same time has been studied in \cite{epl}.

\subsection{Simple Hamiltonian}
The simplest non trivial example that one can consider is that of a two level system, say a spin 1/2, evolving according to a Hamiltonian $H=\omega\, \sigma/2_x$ and that is continuously measured in a basis \emph{different} from the energy basis. Indeed if the measured observable commutes with the Hamiltonian, the system will collapse in one of the Hamiltonian eigenvectors and never jump afterwards, a situation we want to avoid. We thus chose to arbitrarily measure $\mathcal{O}=\sigma_z/2$ at a rate $\gamma^2$ so that the measurement basis is the canonical basis. Eventually we need to rescale the Hamiltonian (or equivalently adimensionalise time) $\omega=\gamma u$ in order to avoid a complete Zeno-freezing of the jumps. In the absence of measurement the system Lindbladian simply reads:
\begin{equation}
\label{eq:simplehamiltonian}
\mathcal{L}(\rho)=-i\frac{\gamma u}{2}[\sigma_x,\rho].
\end{equation}
Expanding equation (\ref{eq:simplehamiltonian}) gives:
\begin{equation}
\left\lbrace
\begin{split}
&\mathcal{A}=0,\\
&\mathcal{B}_0^{10}=\mathcal{B}_1^{01}=-\mathcal{B}_0^{01}=-\mathcal{B}_1^{10}=-iu/2,\\
&\mathcal{C}^0_{01}=\mathcal{C}^1_{10}=-\mathcal{C}^1_{01}=\mathcal{C}^0_{01}=iu/2,
\end{split}\right.
\end{equation}
so that eventually we get the jump rates $m^0_{1}=m^1_0=u^2$. If we reintroduce
the dimension in the Hamiltonian, that is we keep $\omega$ fixed for large yet
fixed $\gamma$ we get an average time between two jumps $\tau=\gamma^2/\omega^2$
which goes to $\infty$ when $\gamma$ goes to infinity. Thus the well known Zeno effect \cite{itano1990}
is recovered in this framework. An analogous result had already been derived in this setting with a less general method in \cite{jordan2013,oqrw,qbm}.
\begin{figure}
\centering
\includegraphics[width=0.48\textwidth]{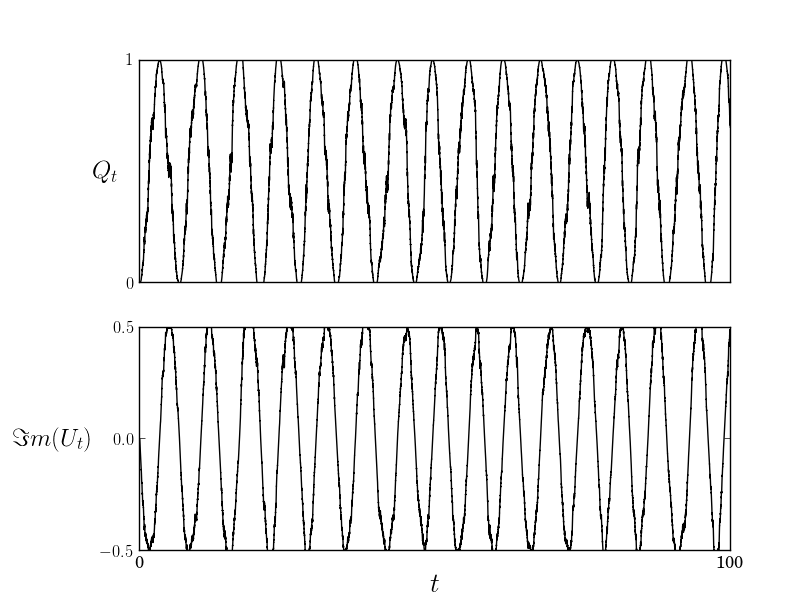}
\includegraphics[width=0.48\textwidth]{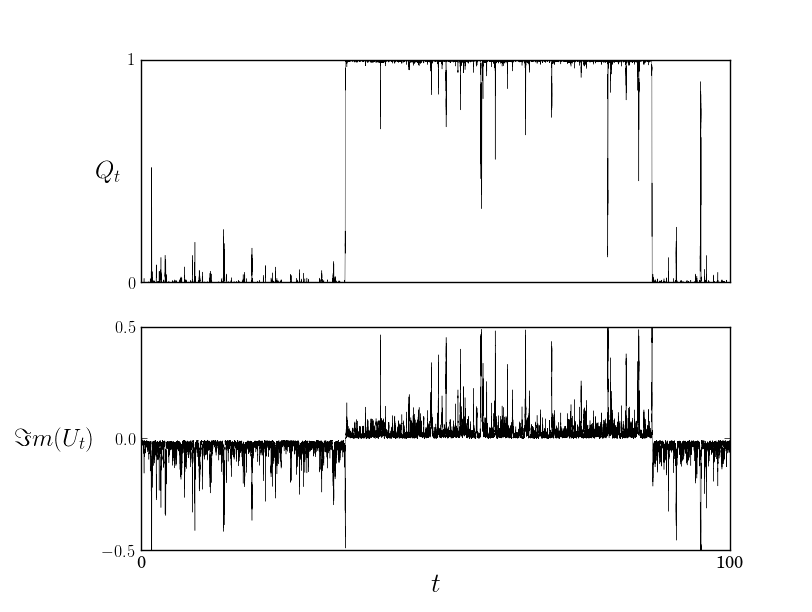}
\caption{Sample trajectory of the density matrix coefficients $Q_t=\bra{1} \rho_t \ket{1}$ and $U_t=\bra{0} \rho_t \ket{1}$ for $\omega=1$ and $\gamma=0.25$ on the left and $\gamma=5.0$ on the right. Notice the change of regime from smooth Rabi oscillations to sharp quantum jumps when $\gamma$ is increased. The remaining thin fluctuations around the limiting jump process are \emph{not} described in our framework.}
\label{fig:simplehamiltonian}
\end{figure}

\subsection{Simple thermal jumps}\label{sq:applicationthermal}
Another application of our result is the study of thermal jumps. The jump rates and more details of the stochastic process had already been derived in \cite{lmp} albeit with an ad-hoc method. We consider a simple two level system consisting of a ground state $\ket{0}$ and an excited state $\ket{1}$ coupled to a thermal bath. It evolves according to an Hamiltonian $H=\frac{\omega}{2}\sigma_z$ and the dissipative part of the evolution is induced by $\sigma_+$ and $\sigma_-$ in the form of two Lindblad generators: \[L_{\sigma_{\pm}}=\sigma_\pm \rho \sigma_\mp - \frac{1}{2} \lbrace\sigma_\mp\sigma_\pm,\rho\rbrace.\]
Eventually the Lindbladian in the absence of measurement reads:
\begin{equation}
\label{eq:simplethermal}
\mathcal{L}(\rho)=-i[H,\rho] + \lambda p L_{\sigma_-}(\rho) + \lambda (1-p) L_{\sigma_+}(\rho),
\end{equation}
where $\lambda$ is the coupling strength with the bath and $p:=1/(1+e^{-\beta \omega})$ is the probability to be in the ground state at thermal equilibrium. This is a simple yet legitimate model for the evolution of a system coupled to a thermal bath \cite{petruccione2002}.
We also assume that its energy is continuously measured (i.e. $\mathcal{O}=\sigma_z$) at a rate $\gamma^2$ and want to characterise the thermally activated quantum jumps that appear in the large $\gamma$ limit. This is actually a trivial question with the help of our proposition. Expanding equation (\ref{eq:simplethermal}) gives, with the previous notations:
\begin{equation}
\left\lbrace
\begin{split}
&\mathcal{A}^1_0=\lambda p, \;\,\mathcal{A}^0_1=\lambda (1-p)\\
&\mathcal{B}=0\\
&\mathcal{C}=0
\end{split}\right.
\end{equation}
So that $\mathcal{A}$ immediately encodes the jump rates. We recover the Gibbs equilibrium distribution for the occupation ratios and see that the jump rates are directly proportional to the system-bath coupling strength.
\begin{figure}
\centering
\includegraphics[width=0.48\textwidth]{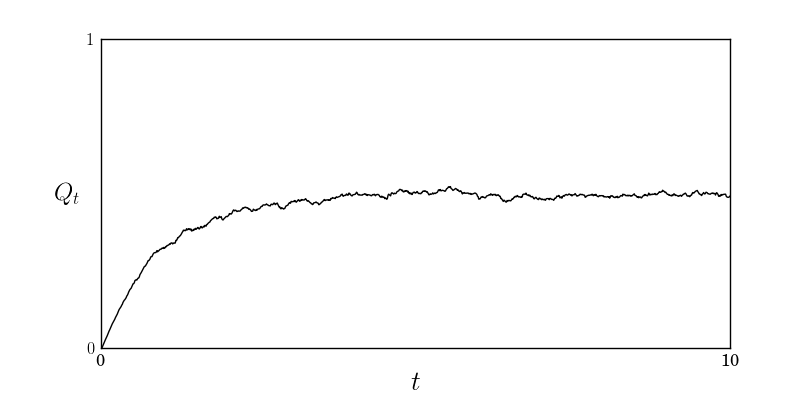}
\includegraphics[width=0.48\textwidth]{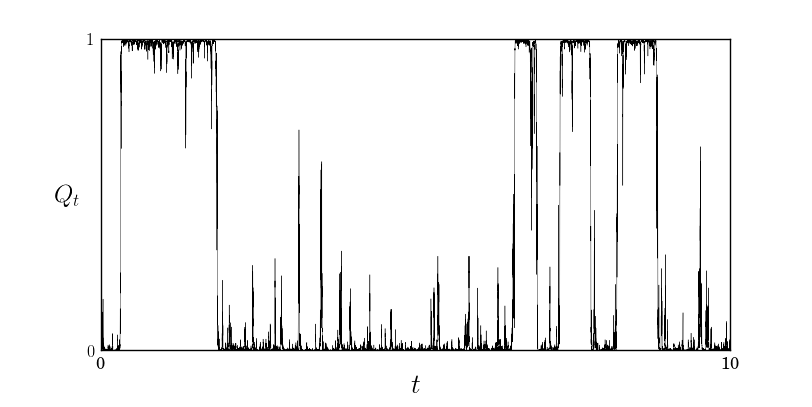}
\caption{Sample trajectory of the density matrix coefficient $Q_t=\bra{1} \rho_t \ket{1}$ with $p=0.5$ for $\gamma=0.05$ on the left and $\gamma=10.0$ on the right. Notice the change of regime from a smooth thermalization to sharp quantum jumps when $\gamma$ is increased.}
\label{fig:simplethermal}
\end{figure}

\section{Proofs} \label{sec:details}

\subsection{Proof of the proposition}\label{proof1}
Here our objective is to prove that a density matrix obeying equation (\ref{eq:main}) becomes a finite state Markov process on the projectors on the eigenvectors of the measured observable in the large $\gamma$ limit and to compute the jump rates $m^i_j$ or equivalently the Markov transition matrix $M$. Using the previous notations equation (\ref{eq:main}) can be expanded to: 
\begin{equation}
\left\lbrace
\begin{split}
dQ_i&=\mathcal{A}({\bf Q})_i\, dt + \mathcal{B}({\bf Y})_i\, dt \\&+ \gamma \sqrt{\eta}\,Q_i\Big(\lambda_i- \sum_k Q_k \lambda_k\Big) dW_t, \\
dY_{ij}&=\gamma^2\Big( \mathcal{C}({\bf Q})_{ij} - \Delta_{ij}\, Y_{ij}\Big) dt \\&+ \gamma \sqrt{\eta}\, Y_{ij} \Big(\nu_i + \bar\nu_j - \sum_k Q_k \lambda_k\Big) dW_t.
\end{split}\right.
\label{eq:scaling}
\end{equation}
We have used the additional notation $Y_{ij}:=\gamma U_{ij}$ and $\Delta_{ij}:= \frac{1}{2}(|\nu_i|^2+|\nu_j|^2-2\nu_i\bar\nu_j)^2+d_{ij}$. Recall that $\lambda_k=\nu_k+\bar\nu_k$. The process is written in terms of the rescaled variables ${\bf Q}$ and ${\bf Y}$ but we shall sometimes abbreviate the notation by using $\rho$ to collectively refer to these variables; for instance $f(\rho)$ is going to be a short name for $f({\bf Q}, {\bf Y})$.

\subsubsection{Strategy:}

The main object we will consider is the probability kernel $K_t(\rho_0,d\rho)$ to go from a given density matrix $\rho_0$ to another density matrix $\rho$, up to $d\rho$, after a time $t$.  The kernel $K_t$ verifies the Kolmogorov equation $\partial_t K_t=K_t\mathfrak{D}$ where $\mathfrak{D}$ is the second order differential operator (or Dynkin operator) associated to the SDE (\ref{eq:scaling}). These concepts will be defined in more details below. At this stage, we do not need to write $\mathfrak{D}$ explicitly but simply to notice that, because of It\^o's formula, the coefficients in front of the noise terms will come squared so that $\mathfrak{D}$ will only contain terms of order 0 and 2 in $\gamma$. As a result we write 
\[ \mathfrak{D}=\mathfrak{D}_0 + \gamma^2 \mathfrak{D}_2,\] 
and will compute $K_t=e^{t\mathfrak{D}_0 + t\gamma^2 \mathfrak{D}_2}$ for large $\gamma$.

The main argument is then the following. Any second order operator associated to well defined SDE's is a non-positive operator, so that when $\gamma$ is very large, even after a small amount of time, only the eigenvectors that are in the kernel of $\mathfrak{D}_2$ will remain when considering $e^{t\gamma^2 \mathfrak{D}_2}$. The idea is then to perform a perturbative expansion around those remaining eigenvectors and compute the jump rate between them.

\subsubsection{Definitions }
We first start by introducing some definitions and notations (see \cite{feller2008,oksendal1992} for more details).
The kernel $K_t$ codes for the probability of a solution of the flow equation started at $\rho_0$ to be at $\rho$ at time $t$. {It can} be used to compute the average of any regular function $f$:
\[ \mathbb{E}[f(\rho_t)] = \int_{\rho\in \mathcal{K}}  K_t(\rho_0,d\rho)\, f(\rho),\]
with $\rho_t$ the solution of the SDE started at $\rho_0$ at time $0$ and the integration domain is the set\footnote{By the Lindblad construction, the flow associated to the SDE defined above preserves this set.} of density operators $\mathcal{K}$. The probability kernel $K_t$ can also be viewed as an operator acting on functions defined on $\mathcal{K}$ via:
\[ K_t: f \to K_tf\;\; \mathrm{with}\;\; (K_tf)(\rho):= \int_{\rho'\in \mathcal{K}}  K_t(\rho,d\rho')f(\rho').\]

An equation {for $K_t$ can} be obtained by computing the time derivative of $\mathbb{E}[f(\rho_t)]$ in two different ways. One can first apply It\^o calculus to $f(\rho_t)$:
\[ df(\rho_t) = (\mathfrak{D}f)(\rho_t)\, dt + (\cdots) dW_t,\]
with $\mathfrak{D}$ the second order differential operator associated to the SDEs (\ref{eq:scaling}) (also called Dynkin operator or sometimes dual Fokker-Planck operator). 
{A function $f(\rho_t)$ is called a local martingale if there is no drift terms in its It\^o derivative, that is if $\mathfrak{D}f=0$.} 

Thanks to the defining property of the Ito calculus, the above equation implies
\[ \partial_t \mathbb{E}[f(\rho_t)] = \mathbb{E}[(\mathfrak{D}f)(\rho_t)] ,\]
Writing the expectations in terms of the  probability kernel $K_t$ gives 
\[ \int_{\rho\in \mathcal{K}}  \partial_t K_t(\rho_0,d\rho)\, f(\rho) =\int_{\rho\in \mathcal{K}}  K_t(\rho_0,d\rho)\, (\mathfrak{D}f)(\rho),\]
for any function $f$. Hence, we get what we had claimed in the introduction of the proof,
\[ \partial_t K_t=K_t\,\mathfrak{D},\]
as an equation on operators acting on functions over $\mathcal{K}$. 
The formal solution of this differential equation, with initial data ${K}_{t=0}=Id$, is 
\[ K_t = \exp(t\mathfrak{D}),\]
again viewed as an operator identity.

The same can be done in the dual picture. If $f(\rho)$  and $\mu(d\rho)$ respectively denote a function and a measure on $\mathcal{K}$, the duality is the obvious one:
\[ <\mu,f>:= \int_{\rho\in \mathcal{K}} \mu(d\rho) \, f(\rho).\] 
If $\mathfrak{O}$ is {an operator} acting on functions, its dual $\mathfrak{O}^T$ acts on measures via
{$<\mathfrak{O}^T\mu,f>:= <\mu,\mathfrak{O}f>$.}
In particular, the dual $\mathfrak{D}^T$ of the Dynkin operator   $\mathfrak{D}$ is the usual Fokker-Planck operator.

By duality, the flow on $\rho$ defined by the SDE induces a flow on the measures via 
\[ \mu_0 \to \mu_t\quad \mathrm{with}\quad  \mu_t(d\rho):=\int_{\rho_0\in \mathcal{K}}  \mu_0(d\rho_0)\, K_t(\rho_0,d\rho),\]
Of course it is such that $<\mu_t, f>=<\mu_0,K_tf>$, so that we can write this flow as $\mu_t=K_t^T\, \mu_0$. By definition we then have
\[ \partial_t\mu_t = \mathfrak{D}^T\,\mu_t,\]
or equivalently $\partial_t K_t^T= \mathfrak{D}^T\, K_t^T$.
A measure is said to be \emph{invariant} if it is constant in time, i.e. if it is annihilated by $\mathfrak{D}^T$. Invariant measures and local martingales are thus dual objects.
Now that we have reminded the reader of these definitions, we can apply the announced strategy.

\subsubsection{The large $\gamma$ limit of the transition kernel}

Let us proceed. Let $F_I$  be the basis diagonalizing the operator $\mathfrak{D}$, let $E_I$ be the associated eigenvalues, $\mathfrak{D}F_I=E_I\, F_I$, $E_I\leq 0$, and let $\Lambda^I$ be the associated dual basis
so that:
\begin{equation} \label{K-expand}
 K_t(\rho_0,d\rho)= \sum_I e^{tE_I}\, F_I(\rho_0)\, \Lambda^I(d\rho).
 \end{equation}

All $E_I$, $F_I$ and $\Lambda^I$ depend on $\gamma$. Since $\mathfrak{D}=\gamma^2\, \mathfrak{D}_2 + \mathfrak{D}_0$, perturbation theory tells that 
\begin{equation*}
 \begin{split}
  E_I&=\gamma^2\,e_I+ e^0_I + \gamma^{-2}\, e_I^1+ O(\gamma^{-4}),\\
  F_I&= f_I + \gamma^{-2} f^1_I + O(\gamma^{-4}),\\
  \Lambda^I&=\mu^I + \gamma^{-2} \mu^I_1 + O(\gamma^{-4}),
 \end{split}
\end{equation*}
with $e_I$ an eigenvalue of $\mathfrak{D}_2$ and $f_I$ (resp. $\mu^I$) the corresponding eigenvector (resp. dual eigenvector). Hence, only the terms  corresponding to eigenvalues $E_I$ whose dominating contribution $e_I$ vanishes
survive in the pointwise limit of the sum (\ref{K-expand}):
\[
\lim_{\gamma\to \infty} K_t(\rho_0,d\rho) = \sum_{I:\, e_I=0} e^{t e_I^0} f_I(\rho_0)\, \mu^I(d\rho).
\] 
Restricting the sum to $e_I=0$ selects $f_I$ to be in the kernel of $\mathfrak{D}_2$: $\mathfrak{D}_2f_I=0$. 


Let now $f_i$ be a basis of $\mathrm{Ker}(\mathfrak{D}_2)$ and $\mu^i$ be the associated dual basis of $\mathrm{Ker}(\mathfrak{D}_2^T)$, so that $f_I$ (with $\mathfrak{D}_2f_I=0$) is a linear combination of $f_i$: $f_I= \sum_j f_j (\int \mu^j f_I)$. 

Degenerate perturbation theory around the zero eigenvalue tells us that $\mathfrak{D}_2 f_I^1 + \mathfrak{D}_0 f_I = e_I^0 f_I$. Integrating this last equation against $\mu_j$ yields
\[ 
\int \mu^j(d\rho)(\mathfrak{D}_0 f_I)(\rho) = e_I^0\, \int \mu^j(d\rho) f_I(\rho),
\]
so that $e_I^0$ are eigenvalues of the matrix $m^j_k:= \int \mu^j\,\mathfrak{D}_0 f_k$ and $f_I$ the corresponding eigenvectors.

Consequently, we may write the limiting formula for the point-wise limit of the kernel $K_t$ in the basis $f_i$ as
\begin{equation} \label{K-limit}
\lim_{\gamma\to \infty} K_t(\rho_0,d\rho) = \sum_{ij} f_j(\rho_0)\, \left({ e^{tM} }\right)^j_i\, \mu^i(d\rho),
\end{equation}
with
\[
m^j_i := \int \mu^j(d\rho)\, (\mathfrak{D}_0 f_i)(\rho),
\] 
where $f_i$ is a basis of
$\mathrm{Ker}\mathfrak{D}_2$ and $\mu^j$ is the dual basis. Recall that
$\mathfrak{D}_2$ is the second order differential operator associated to SDE's.
Hence, to any element of $\mathrm{Ker} \mathfrak{D}_2$ corresponds a (local)
martingale for the associated stochastic process. 
{The dual basis of invariant measures $\mu^i \in \mathrm{Ker} \mathfrak{D}_2^T$ that we shall identify below have disjoint supports so that we can use those to index the states of a finite dimensional Markov chain (we say that the chain is in the state $i$ if $\rho$ is in the support of $\mu^i$) and to ensure the consistency of the associated process.}
The large $\gamma$ limit thus
projects the original process on a finite state Markov  {chain} whose states are
indexed by the $\mathfrak{D}_2$-martingales and the matrix $M$ contains the
transition probabilities between the states.

\subsubsection{Computing the Markov transition matrix:}

We now need to compute the $f$'s and the $\mu$'s to eventually get the Markov matrix $M$. The operator $\mathfrak{D}_2$ is the second order operator associated to the set of SDE's:
\begin{equation}
\left\lbrace
\begin{split}
dQ_i&= \sqrt{\eta}\,Q_i\Big(\lambda_i- \sum_k Q_k \lambda_k\Big) dW_t \\
dY_{ij}&=\Big( \mathcal{C}(\mathbf{Q})_{ij} - \Delta_{ij}\, Y_{ij}\Big) dt \\&+\sqrt{\eta}\, Y_{ij} \Big(\nu_i + \bar\nu_j - \sum_k Q_k \lambda_k\Big) dW_t
\end{split}\right.
\label{eq:D2}
\end{equation}
To compute the invariant measures associated to this SDE, we could try to solve its Fokker-Planck equation. However as the invariant measures are singular in the $Q$'s, it is easier to study the stochastic process directly. Let us first notice that the $Q$'s are bounded martingales and as a result they converge almost surely to one of the fixed points. It is easy to see that the only fixed points of the noise term are of the form $Q_k=\delta_{ik}$ where $i$ is random and depends on the trajectory. (Recall that we assumed that all $\lambda_k$ were distinct). This is an expected result because eqs.(\ref{eq:D2}), but without the $\mathcal{C}(\mathbf{Q})_{ij}$-terms, are those for indirect non-demolition measurements: it simply means that, because of measurement, the probabilities tend to collapse with all the mass in a (random) {pointer state}. These are all  linearly independent martingales and 
\[ f_i(\rho):= Q_i \in \mathrm{Ker}\mathfrak{D}_2.\]
form a basis of $\mathrm{Ker}\mathfrak{D}_2$.  Notice that $\sum_i f_i=1$ and this ensures that $\sum_i m^j_i=0$. There are as many dual invariant forms $d\mu$'s as there are fixed points and we have 
\[ \mu^i(d\rho):=\delta^i(d\mathbf{Q})\, \mu^i(d\mathbf{Y}) \in \mathrm{Ker}\mathfrak{D}_2^T,\]
where $\delta^i(d\mathbf{Q})=\delta(1- Q_i) dQ_i \prod_{j\neq i} \delta(Q_j) dQj$, with $\delta(Q)dQ$ the Dirac measure, and $\mu^i(d\mathbf{Y})$ is the normalized $Y$-dependence of the invariant measure conditioned on the fact that the trajectory in the $Q$-subspace converge to $Q_j=\delta_{ij}$. Notice that in general, $\mu^i(d\mathbf{Y})$ has no reason to be peaked and is actually rather broad with power law tails.  Computing $\mu^i(d\mathbf{Y})$ is difficult in general because $\mathcal{C}^k_{ij}$ need not be real. 
A possible solution is to solve the equation for $Y_{kl}$ to compute its moments.
But it is even easier to notice that neither $\mu^i(d\mathbf{Y})$ nor all its moments will actually ever be needed to compute the transition rates $\int \mu^i \mathfrak{D}_0 f_j$ which only depend on the average value of $\mathbf{Y}$. This average can be easily computed. Let us  integrate the SDE for $Y_{kl}$ conditioned on $Q_l\rightarrow\delta_{il}$ for all $l$:
\begin{equation}
Y_{kl}=\int_0^t\Big( \mathcal{C}^i_{kl} - \Delta_{kl} Y_{kl}\Big) dt +\sqrt{\eta}\int_0^t Y_{kl} \Big(\nu_k + \bar\nu_l - \lambda_i\Big) dW_t
\end{equation}
We now take the average and write $y^i_{kl}:=\mathds{E}[Y_{kl}| {\bf Q_\cdot}=\delta_{\cdot i}]$ which gives $\dot y^i_{kl}=\mathcal{C}^i_{kl}-\Delta_{kl} y^i_{kl}$ so that $y^i_{kl}(t)=\frac{C_{kl}^i}{\Delta_{kl}} + y^i_{kl}(0)e^{-\Delta_{kl} t}$, and, since $\re\, \Delta_{kl}>0$,
\[ y^i_{kl}(t):= \mathds{E}[Y_{kl}|{\bf Q_\cdot}=\delta_{\cdot i}]\underset{t\rightarrow+\infty}{\longrightarrow} \frac{C_{kl}^i}{\Delta_{kl}}.\]
Recall that $\Delta_{kl}:= \frac{1}{2}(|\nu_k|^2+|\nu_l|^2-2\nu_k\bar\nu_l)^2+d_{kl}$.

We can now compute the transition rates $m^i_j=\int \mu^i \mathfrak{D}_0f_j$.
The operator $\mathfrak{D}_0$ can be easily computed as it is the operator associated to the (S)DE $\upd Q_i=\mathcal{A}(\mathbf{Q})_i \upd t + \mathcal{B}(\mathbf{Y})_i \upd t$, without noisy terms, so that $\mathfrak{D}_0$ is the first order differential operator:
\begin{equation}
\mathfrak{D}_0=\sum_i \left[\mathcal{A}(\mathbf{Q})_i + \mathcal{B}(\mathbf{Y})_i \right] \partial_{Q_i}
\end{equation}
Recalling that $f_j(\rho)=Q_j$ we get (with implicit summation on repeated indices):
\[
\begin{split}
m^i_j&=\int \mu^i \mathfrak{D}_0 f_j\\
&=\int_{\mathcal{K}} \delta^i(d\mathbf{Q})\, \mu^i(d\mathbf{Y})\, \left(\mathcal{A}^k_j Q_k+\mathcal{B}^{kl}_j Y_{kl}\right)\\
&=\mathcal{A}^k_j \int_{\mathcal{K}} \delta^i(d\mathbf{Q})  Q_k \mu^i(d\mathbf{Y}) + \mathcal{B}^{kl}_j \int_{\mathcal{K}} \mu^i(d\mathbf{Y}) Y_{kl}\delta^i(d\mathbf{Q})\\
&=\mathcal{A}^i_j + \sum_{kl} \frac{\mathcal{B}^{kl}_j \mathcal{C}_{kl}^i}{\Delta_{kl}},
\end{split}
\]
which is the result that was announced previously.

\subsection{What is the most general jumpy scaling limit?}\label{proof:scaling}
We now prove that we have derived the most general scaling limit \emph{that gives rise to quantum jumps}. For $\mathcal{A}$, $\mathcal{B}$ and $\mathcal{C}$ we obviously have the most general scaling as if they were scaled with a smaller power, they would become irrelevant in the scaling limit (or equivalently the Zeno effect would kill the associated transition rates), and if they were scaled with a bigger power, the jump rates would diverge in the large $\gamma$ limit and the limit would not be jumpy anymore. To say it differently, the simple fact that we ask the limit to be jumpy and that the system parameters have an influence fixes completely the scaling. We now need to show, as we have previously announced, that the phase-phase coupling term $D$ in the Lindbladian is always irrelevant no matter how it is rescaled \emph{unless it is diagonal}. We will first show that that the non diagonal terms cannot grow faster than $\gamma$ and then prove that such a limiting scaling would still make them irrelevant. Eventually we prove the link between the scaling expressed in terms of $A$, $B$, $C$ and $D$ and the Linblad generators of remark \ref{rem}.
\setcounter{paragraph}{0} 
\paragraph{D cannot grow faster than $\gamma$, unless diagonal:}
Let us suppose that $D$ can grow faster than $\gamma$. It would then need to be rescaled independently of $A$, $B$ and $C$ which were shown to grow no faster than $\gamma$. As a result $D$ needs to be itself the generator of a completely positive map $\Phi_t$ that couples only the phases with the phases.
Let us suppose that we have a completely positive application $\Phi_t$ with generator $\tilde{ \mathcal{L}}$ that couples only the phases and write it using the usual decomposition:
\begin{equation}
\tilde{\mathcal{L}}(\rho)= -i\big[H,\rho\big]+ \sum_a\Big( M^a\rho M^{a\, \dagger} - \frac{1}{2}\{M^{a\, \dagger} M^a,\rho\}\Big),
\label{eq:lindblad}
\end{equation}
for some operators $M^a$.
We ask that $\tilde{\mathcal{L}}$ does not act on the diagonal coefficients, i.e. $\tilde{\mathcal{L}}(\ket{i} \bra{i})=0$ for any projector on the measurement eigenvector $\ket{i} \bra{i}$. In particular, imposing that the diagonal elements vanish, i.e. $\bra{j}\tilde{\mathcal{L}}(\ket{i} \bra{i})\ket{j}=0$ for any $j$, reads
\[
\sum_a\Big( \big\vert{M^a_{ji}}\big\vert^2-\delta_{ij} (M^{a\, \dagger} M^a)_{jj}\Big)=0.
\]
Thus, for $j \neq i$, we get 
\begin{equation}
\sum_a \big\vert M^{a}_{ji}\big\vert^2=0,
\end{equation}
and hence $M^a_{ji}=0$ for all $a$ and $j\not= i$. 
That is, all the $M^{a}$'s are diagonal matrices. The Hamiltonian part of the flow also needs to be diagonal. Indeed, if $H_{kl} \neq 0$ for some $k$ and $l$, $k\neq l$, we have $\bra{k}[H,\ket{l}\bra{l}]\ket{l}=H_{kl}$ so $H$ couples the probabilities to the phase which is forbidden. Therefore, $H$ is also diagonal and as a result $\tilde{\mathcal{L}}$ cannot mix the phases. Writing $M^a=\sum_k n_k^a\ket{k}\bra{k}$, this means that at most $D(\mathbf{Y})_{ij}=-D_{ij} Y_{ij}$, with $D_{ij}=\frac{1}{2}\sum_a(|n_i^a|^2+|n_j^a|^2-2n_i^a \bar n_j^a)+i (H_{ii}-H_{jj})$, a term that is proportional to the deterministic part of the measurement acting on the phase. The only non-trivial consistent scaling is that $D$ scales as $\gamma^2$ and $D(Y)_{ij}= - \gamma^2 d_{ij} Y_{ij}$. Notice that we have proved at the same time that in the generators picture, the terms contributing to $D$ come from a diagonal Hamiltonian and the diagonal part of the matrices $M^a$ appearing in the Lindblad generator.

\paragraph{If D scales as $\gamma$ then it is irrelevant:}
Let us suppose that $D=\gamma \mathcal{D}$ which is the limiting scaling allowed for a non-diagonal $D$. In that case the Fokker-Planck operator associated to equation (\ref{eq:main}) needs to be written with a new term of order $\gamma$ that is $\mathfrak{D}=\mathfrak{D}_0+\gamma \mathfrak{D}_1 + \gamma^2\mathfrak{D}_2$ where $\mathfrak{D}_1$ is the Fokker-Planck operator associated to the (S)DE: $\upd Y_{ij}= \mathcal{D}^{kl}_{ij} Y_{kl}\, dt$. We thus have:
\[\mathfrak{D}_1=\sum_{ijkl} \mathcal{D}^{kl}_{ij} Y_{kl}\partial_{Y_{ij}}\]

We now proceed with the same perturbative expansion as before except this time we assume that a term of order $\gamma$ remains, i.e. that $K_t= e^{t\gamma^2\mathfrak{D}_2 + t \gamma \mathfrak{D}_1 + t\mathfrak{D}_0}$. The first two terms of the eigenvalue expansion will need to be zero to give a non trivial jumpy behavior in the large $\gamma$ limit. We use the same notation as before, that is:
\begin{equation}
 \begin{split}
  F_I&=f_I+\gamma^{-1}f_I^1+\gamma^{-2}f_I^2 + ...\\
  E_I&=0 \times \gamma^2 + 0 \times \gamma + e^0_I + ...
 \end{split}
\end{equation}
for the eigen-modes with leading vanishing eigenvalues.
We then have, up to second order in $\gamma$:
\begin{equation}
 \begin{split}
  &\mathfrak{D}_2 f_I=0\\
  &\mathfrak{D}_2 f_I^1 + \mathfrak{D}_1 f_I = 0\\
  &\mathfrak{D}_2 f_I^2 + \mathfrak{D}_1 f^1_I + \mathfrak{D}_0 f_I = e_I\, f_I\\
  \end{split}
\end{equation}
Recall that the $f$'s only depend on $\mathfrak{D}_2$ and are thus the same as before, that is $f_i(\mathbf{Q},\mathbf{Y})=Q_i$ which gives in particular $\mathfrak{D}_1 f_I=0$. In that case we have $\mathfrak{D}_2 f_I^1=0$ so that $f_I^1 \in \mathrm{Ker}(\mathfrak{D}_2)$ and as a consequence, $\mathfrak{D}_1 f_I^1$ is also zero. As a result, the terms of order $\gamma$ no longer have any role to play in the computations and $\mathcal{D}$ is irrelevant. 
 
\paragraph{The scaling provided in remark 3 is the most general:}
Now that we have the most general scaling in terms of $A$, $B$, $C$ and $D$, we only need to relate them to the expression of the Lindblad generators. Using the generic Linbladian of equation (\ref{eq:lindblad}) we get:
\[
A^i_j=\sum_a\Big( \big\vert{M^a_{ji}}\big\vert^2-\delta_{ij} (M^{a\, \dagger} M^a)_{jj}\Big)
\]
This show that if the $M_a$'s have non diagonal parts, then they have to be of order zero in $\gamma$. As a result, the contribution of the $M_a$'s having non diagonal parts to $B$ and $C$ is also of order zero and vanishes once we consider the rescaled coefficients $\mathcal{B}$ and $\mathcal{C}$. If the $M_a$'s are purely diagonal, we have already proved that they only contribute to $D$. As a result, the only contributions to $\mathcal{B}$ and $\mathcal{C}$ come from the Hamiltonian and we compute:
\[
{C}^i_{kl}{B}^{kl}_{j}=\left(H_{il}\delta_{ik}-H_{ki}\delta_{il}\right)\left(H_{jk}\delta_{jl}-H_{lj}\delta_{jk}\right)
\]
This means that only the non diagonal parts contribute as $k\neq l$ and they need to be rescaled as $\gamma$ to be relevant. Eventually, the diagonal parts of the $M_a$'s and of $H$ appear in $D$ and consequently need to be rescaled with a factor $\gamma^2$ to stay relevant in the large $\gamma$ limit. To summarise, the non diagonal coefficients of the $M_a$'s need not be rescaled, the non diagonal coefficients of the Hamiltonian need to be rescaled with a factor $\gamma$ and the diagonal part of the Hamiltonian and the $M_a$'s need to be rescaled with a factor $\gamma^2$. This proves the form of equation (\ref{eq:generatorscaling}).

\section{Discussion}\label{sec:discussion}
After these lengthy derivations, let us step back and comment on the mathematical results and their physical implications. We have shown that an open quantum system that is continuously measured has an evolution that gets jumpy when the measurement process dominates. Our derivation shows that quantum jumps are ubiquitous in the sense that any quantum system subjected to a tight monitoring will undergo quantum jumps (or will simply be frozen, i.e. will have a jump rate equal to zero).

However, we should insist that for large but finite $\gamma$, the evolution of the system density matrix is still continuous and the jumps --though they look instantaneous in the limit-- have a finite duration of order $\gamma^{-2}$. In this setting, quantum jumps are not strictly instantaneous and are only the effective description of a more fundamental evolution. This is in stark contrast with the quantum jumps that appear directly in stochastic master equations for Poissonian unravellings. Our derivation thus gives some insights into  the debate about the reality of quantum jumps. The conclusions will however depend on the foundational attitude of the reader. If one is ready to give an ontological status to the conditioned density matrix -- as it is the case for example in dynamical reduction models \cite{ghirardi1986,bassi2003}--, then the jumps we observe are really a consequence of measurement and can even be assumed to be \emph{progressively created} by measurement. From a more epistemic or Bayesian perspective, measurements could just be progressively revealing, as well as influencing, a (yet to be specified) underlying jump process. As far as we know, a specific model for the second option has not been provided yet --though it should certainly be investigated. 

From a more practical point of view, we have provided a simple analytical recipe to compute the effective evolution of systems that are either continuously monitored or subjected to strong dissipation. The jump rates are simple analytical functions of the measured operator and the Linbladian. Our results have been derived for a single measured observable but can effortlessly be generalised to a larger set of commuting observables. Finally, we should add that we have voluntarily neglected the study of the remaining fluctuations around the jump process. This admittedly difficult but fascinating enquiry is, we believe, the next step in the thorough understanding of quantum jumps.

\begin{acknowledgments}
This work was supported in part by the ANR contracts ANR-2010-BLANC-0414 and ANR-14-CE25-0003-01.
\end{acknowledgments}

\bibliography{main}
\bibliographystyle{plain}
\end{document}